\documentclass[11pt, onecolumn]{article}
\usepackage{times,bm}
\usepackage[affil-it,]{authblk}
\usepackage[pagebackref=false,colorlinks=true,linkcolor=olive,citecolor=olive,urlcolor=olive]{hyperref}
\usepackage{geometry}
\usepackage[switch*, displaymath,pagewise, mathlines]{lineno}
\usepackage{graphicx}
\usepackage{cite}
\usepackage{amsmath}
\usepackage{amsfonts}
\usepackage{amssymb}
\usepackage{slashed}
\usepackage{subfloat}
\usepackage{slashed}
\usepackage{color}
\usepackage{float}
\usepackage{multirow,multicol}
\usepackage{tabulary}
\usepackage[flushleft]{threeparttable}
\usepackage{pgfplots,subfigure}
\usepackage{xcolor}
\numberwithin{equation}{section}
\usepackage{tikz}
\usepackage{ulem}
\usepackage{hyperref}
\usepackage{nameref}
\usepackage{comment}
\usepackage{xcolor}
\definecolor{acsblue}{RGB}{17,76,139}
\definecolor{shadecolor}{RGB}{255,241,204}

\usepgflibrary{arrows}
\usetikzlibrary{shapes.callouts}
\tikzset{
	level/.style   = { thick, },
	connect/.style = { dotted, red   },
	notice/.style  = { draw, rectangle callout, callout relative pointer={#1} },
	label/.style   = { text width=2cm }
}

\usepackage{letltxmacro}
\makeatletter
\let\oldr@@t\r@@t
\def\r@@t#1#2{%
	\setbox0=\hbox{$\oldr@@t#1{#2\,}$}\dimen0=\ht0
	\advance\dimen0-0.2\ht0
	\setbox2=\hbox{\vrule height\ht0 depth -\dimen0}%
	{\box0\lower0.4pt\box2}}
\LetLtxMacro{\oldsqrt}{\sqrt}
\renewcommand*{\sqrt}[2][\ ]{\oldsqrt[#1]{#2}}
\makeatother
\usepackage{environ}

\begin{document}

\def\nofundrefquery{}
\def\stmdocstextcolor#1{}
\def\stmdocscolor#1{}
\newcommand{{\ri}}{{\rm{i}}}
\newcommand{{\Psibar}}{{\bar{\Psi}}}
\renewcommand{\rmdefault}{ptm}

\title{\mdseries{Minimally coupled fermion-antifermion pairs via exponentially decaying potential}}
\author{ \textit{Abdullah Guvendi}$^{\ 1}$\footnote{\textit{E-mail: abdullah.guvendi@erzurum.edu.tr } }~,~ \textit {Semra Gurtas Dogan}$^{\ 2}$\footnote{\textit{E-mail: semragurtasdogan@hakkari.edu.tr \textcolor{olive}{(Cor. Auth.)}} } ~,~ \textit {Omar Mustafa}$^{\ 3}$\footnote{\textit{E-mail: omar.mustafa@emu.edu.tr} }  \\
	\small \textit {$^{\ 1}$ Department of Basic Sciences, Erzurum Technical University, 25050, Erzurum, Türkiye}\\
 \small \textit {$^{\ 2}$ Department of Medical Imaging Techniques, Hakkari University, 30000, Hakkari, Türkiye}\\
	\small \textit {$^{\ 3}$ Department of Physics, Eastern Mediterranean University, 99628, G. Magusa, north Cyprus, Mersin 10 - Türkiye}\\}

\date{}
\maketitle
\begin{abstract}
In this study, we explore how a fermion-antifermion ($f\overline{f}$) pair interacts via an exponentially decaying potential. Using a covariant one-time two-body Dirac equation, we examine their relative motion in a three-dimensional flat background. Our approach leads to coupled equations governing their behavior, resulting in a general second-order wave equation. Through this, we derive analytical solutions by establishing quantization conditions for pair formation, providing insights into their dynamics. Notably, we find that such interacting $f\overline{f}$ systems are unstable and decay over time, with the decay time depending on the Compton wavelength of the fermions.

\end{abstract}

\begin{small}
\begin{center}
\end{center}
\end{small}


\bigskip

\section{\mdseries{Introduction}}\label{sec1}

In the realm of quantum mechanics, understanding the behavior of $f\overline{f}$ pairs under the influence of various potentials is crucial for unraveling the intricacies of particle interactions. One particularly intriguing scenario arises when these particles interact through exponentially decaying potentials \cite{book1,book2}. This phenomenon not only sheds light on the fundamental nature of fermionic interactions but also holds significance in diverse areas ranging from high-energy particle physics to condensed matter physics. The study of $f\overline{f}$ pairs interacting via exponentially decaying potentials presents a fascinating avenue for exploring the delicate balance between attractive and repulsive forces within quantum systems. Unlike other potentials, which may exhibit linear or polynomial decay, the exponential decay introduces unique characteristics that profoundly influence the behavior of the interacting particles. In this investigation, we will try to understand the underlying principles governing the dynamics of fermionic systems subject to an exponentially decaying potential. Through a theoretical analysis, we aim to elucidate the impact of various parameters on the resulting phenomena, unraveling the rich tapestry of quantum behavior encapsulated within these systems. Furthermore, the insights gleaned from this study are anticipated to have far-reaching implications across multiple disciplines. From elucidating the behavior of quark-antiquark pairs in the context of quantum chromodynamics to shedding light on the emergence of exotic states in condensed matter systems, the ramifications of understanding $f\overline{f}$ interactions mediated by exponentially decaying potentials are manifold. Furthermore, deriving closed-form analytical solutions for the renowned wave equations, especially those involving exponentially decaying potentials \cite{book1,book2}, seems unfeasible across all scenarios. 

Various studies have examined the dynamics of fermions influenced by exponentially decaying interaction potentials. For instance, under the effect of these potentials, de Castro and Hott \cite{EP1} derived analytical solutions for the Dirac equation in $(1+1)$-dimensions. Pe{\~n}a et al. \cite{EP2} reported analytical bound state solutions with spin and pseudo-spin symmetries, using the Green-Aldrich approximation for the centrifugal term. Long and colleagues \cite{EP3} explored the behavior of relativistic fermions in the context of minimum length using the Bethe ansatz method. Ikot \cite{EP4} investigated the Dirac equation considering a generalized Hylleraas potential through the extension of the Nikiforov-Uvarov method. Arda et al. \cite{EP5} found approximate analytical solutions for the pseudo-spin symmetric Dirac equation. Moreover, Dirac fermions have been studied across various scalar interaction potentials, including exponentially decaying potentials (see, e.g., \cite{EP6}). Consequently, approximations are often employed to handle these complex systems involving exponential potentials \cite{book1,book2}. To mitigate complexity, one-body test fields are frequently utilized, though it is important to underline that exponentially decaying potentials are predominantly utilized in the context of mutually interacting particles \cite{book1,book2}. However, the investigation into relativistic $f\overline{f}$ pairs and fermion-fermion systems interacting via exponentially decaying inter-particle interaction potential remains elusive. Moreover, in the theoretical examination of composite systems formed by interacting particles, it is imperative that the equations encompass the kinematics of each particle, as noted by Barut \cite{barut}. 

In classical quantum mechanics, a widely accepted approach for characterizing bound, scattering, and resonance states entails employing one-time equations formulated with wave functions contingent upon the coordinates of individual particles. These equations incorporate independent Hamiltonians for each particle plus inter-particle interaction potentials. However, transitioning to the realm of relativistic physics poses several challenges in elucidating the behavior of many-body systems. Among these hurdles is the "many-time problem," stemming from retardation effects, alongside the intricacies surrounding the total spin of composite systems comprising multiple spinning particles. Consequently, the requisite equations must be one-time, entirely covariant formulations that encompass many bodies, accounting for retardation effects and accurately encapsulating spin algebra \cite{barut}. Moreover, such equations must integrate the most comprehensive electromagnetic potentials, alongside spinor fields contingent upon spacetime position vectors for each particle. To delineate the dynamics of fermionic many-body systems, a non-perturbative equation emerged from quantum electrodynamics by leveraging the action principle \cite{barut}. Additionally, it was established that this equation could be derived as an excited state of Zitterbewegung \cite{barut-2}. Nevertheless, in ($3+1$)-dimensions, exact solutions to this equation remain elusive, even for renowned two-body systems like one-electron atoms \cite{nuri}, positronium-like unstable systems \cite{nuri-2}, and other interacting $f\overline{f}$ pairs \cite{mos}. This is predominantly due to the incorporation of spin algebra in the aforementioned equation, constructed through the Kronecker product of Dirac matrices, resulting in a set of $16$ radial equations governing the relative motion of $ff$ or $f\overline{f}$ systems upon separating the radial and angular components. In $(3+1)$- dimensions, this equation engenders coupled second-order equations, with subsequent studies demonstrating that perturbation methods alone can address these resultant equations \cite{nuri-2,mos}. Nonetheless, indications suggest that obtaining exact solutions of the fully-covariant two-body Dirac equation appears plausible for low-dimensional systems or specific $f\overline{f}$ systems characterized by dynamical symmetry \cite{guvendi1,guvendi2,guvendi3,guvendi4,guvendi5}. This fully-covariant equation has found application in diverse scenarios, ranging from elucidating the dynamics of an  exciton in a monolayer medium \cite{guvendi1}, estimating mass spectra for neutral mesons \cite{guvendi2}. Furthermore, this equation's utility extends to scrutinizing the evolution of certain $ff$ or $f\overline{f}$ systems in curved spaces \cite{guvendi3,guvendi4,guvendi5,semra,guvendi-mustafa}.

In this manuscript, we endeavor to explore the motion of a minimally coupled\footnote{In the context of the Dirac equation, the interaction potential terms can be classified into three main categories: scalar, non-minimal, and minimal. Scalar potentials only affect the time component of the Dirac equation, essentially altering the particle's mass. This alters the energy across all momentum states. Non-minimal potentials couple the fermions to external fields, resulting in additional terms in the Dirac equation that involve the spin operator and the electromagnetic vector potential. Minimal interactions, on the other hand, involve the coupling of the fermions to the electromagnetic field via the electromagnetic vector potential. These potentials arise from the minimal coupling prescription in quantum field theory, where the fermion's charge couple to the electromagnetic field in the most straightforward manner.} $f\overline{f}$ pair within the realm of relativistic quantum mechanics, specifically focusing on their interaction through an exponentially decaying potential. Our approach involves employing the fully-covariant two-body equation. To achieve this, we introduce the two-body Dirac equation within a three-dimensional spacetime background, both globally and locally flat. For any arbitrary inter-particle interaction potential, we derive a set of coupled equations that govern their relative motion. This set of equations leads to a second-order wave equation, encompassing a general central potential. By considering an exponentially decaying potential as the inter-particle interaction potential, we derive a conditionally exact analytical solutions. In doing so, we establish quantization conditions for the formation of such pairs, offering valuable insights into their dynamic behavior.

This paper is structured as follows: In section \ref{sec:2}, we introduce the corresponding two-body equation and derive a matrix equation. In section \ref{sec:3}, we present a conditionally exact solution (owing to a three-term recurrence relation) of the resulting second-order wave equation. Finally, in section \ref{sec:4}, we provide a summary and discuss the fundamental characteristics of such systems.

\section{\mdseries{Matrix equation}}\label{sec:2}

\begin{figure}
\centering
\includegraphics[scale=0.70]{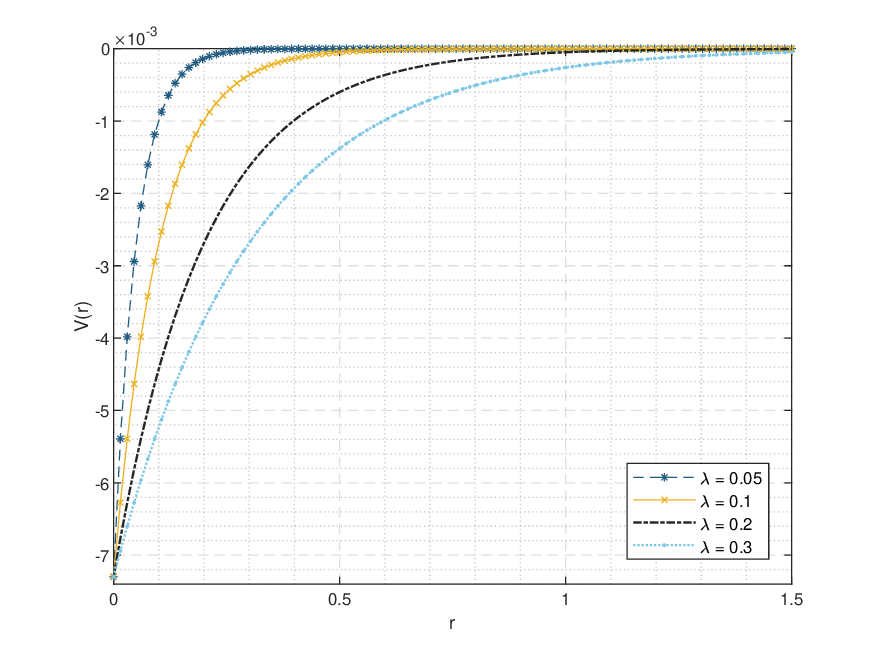}
\caption{Inter-particle interaction potential $\mathcal{V}(r)$ vs $r$. Here, we set $\alpha=1/137$ and $Z=1=d$.}
\label{fig:1}
\end{figure}

In this section, let us start by introducing the fully-covariant two-body Dirac equation in $2+1$-dimensional globally and locally flat spacetime background that can be described by the following metric with the signature $(+,-,-)$
\begin{equation}
ds^2=c^2dt^2-dx^2-dy^2, \label{metric}
\end{equation}
where $c$ is the light speed and $t,x,y$ are the coordinates within the considered flat spacetime. For a mutually interacting fermion-fermion systems in this space-time background, the two-body Dirac equation can be written as the following \cite{guvendi2}
\begin{eqnarray}
&\left\lbrace \left[\gamma^{\mu^{f}}\mathcal{D}^{f}_{\mu}+i\frac{mc}{\hbar}\textbf{I}_2 \right] \otimes\gamma^{t^{\overline{f}}}+\gamma^{t^{f}}\otimes \left[\gamma^{\mu^{\overline{f}}}\mathcal{D}^{\overline{f}}_{\mu}+i\frac{mc}{\hbar}\textbf{I}_2 \right]\right\rbrace \Psi\left(x^{f}_{\mu},x^{\overline{f }}_{\mu}\right)=0,\\
&\mathcal{D}^{f}=\partial_{\mu}^{f}+ie^{f}\frac{\mathcal{A}_{\mu}^{\overline{f}}}{\hbar c},\quad \mathcal{D}^{\overline{f}}=\partial_{\mu}^{\overline{f}}+ie^{\overline{f}}\frac{\mathcal{A}_{\mu}^{f}}{\hbar c}, \label{eq1}
\end{eqnarray}
for a $f\overline{f}$ ($f\overline{f}$) pair with mass of $m$ (each). Here, $e$ stands for electric charge, $\hbar$ is the reduced Planck constant, $\mathcal{A}_{\mu}$ is the electromagnetic 3-vector potential, the Greek indices indicate the coordinates within the considered background and the symbol $\otimes$ means Kronecker product. Here, $\Psi$ is the bi-local spinor field dependent on the space-time position vectors ($x^{f}_{\mu},x^{\overline{f }}_{\mu}$) of the particles and it is constructed through direct product of symmetric two Dirac spinors. The generalized Dirac matrices ($\gamma^{\mu}$) are determined by using the flat Dirac matrices ($\overline{\gamma}^{a}$, $a=0,1,2.$) and inverse tetrad fields ($e^{\mu}_{a}$) through the relation: $\gamma^{\mu}=e^{\mu}_{\,a}\,\overline{\gamma}^{a}$. The flat Dirac matrices are chosen in terms of Pauli spin matrices as the following: $\overline{\gamma}^{0}=\sigma_{z}$, $\overline{\gamma}^{1}=i\sigma_{x}$ and $\overline{\gamma}^{2}=i\sigma_{y}$ according to the signature ($+,-,-$) since $\sigma_{x(y,z)}^{2}=\textbf{I}_{2}$ where $\textbf{I}_{2}$ is two-dimensional identity matrix. The inverse tetrad fields are determined through $e^{\mu}_{\,a}=g^{\mu\nu}e_{\, \nu}^{b}\eta_{ab}$ where $g^{\mu\nu}=\textrm{diag}(c^{-2},-1,-1)$ is the contravariant metric tensor and $\eta_{ab}=\textrm{diag}(1,-1,-1)$ is the flat Minkowski tensor. The tetrads can be obtained by the relation: $g_{\mu\nu}=e_{\,\mu}^{a}e_{\,\nu}^{b}\eta_{ab}$ where $g_{\mu\nu}=\textrm{diag}(c^2,-1,-1)$ is the covariant metric tensor. Accordingly, we can obtain the generalized Dirac matrices as the following
\begin{equation}
 \gamma^{t^{f(\overline{f})}}=\frac{\sigma_{z}}{c},\quad  \gamma^{x^{f(\overline{f})}}=i\sigma_{x},\quad  \gamma^{y^{f(\overline{f})}}=i\sigma_{y},  \label{GDM}
\end{equation}
where $i=\sqrt{-1}$. Here, we are interested in a $f\overline{f}$ pair interacting through a central interaction potential without considering any other external force. Hence, we can take into account this interaction through the following components of the electromagnetic 3-vector potential
\begin{equation*}
\mathcal{A}_{t}=\mathcal{V}(|x^{f}_{\mu}-x^{\overline{f }}_{\mu}|),\quad \mathcal{A}_{x}=0=\mathcal{A}_{y}.
\end{equation*}
Now, it can be useful to write the corresponding matrix equation in the following form $\hat{\diamondsuit}\Psi=0$, where $\hat{\diamondsuit}$ is
\begin{eqnarray}
&\gamma^{t^{f}}\otimes\gamma^{t^{\overline{f}}}\left[\partial_{t}^{f}+\partial_{t}^{\overline{f}} +i\mathcal{V}\right]+\gamma^{x^{f}}\partial_{x}^{f}\otimes \gamma^{t^{\overline{f}}}+ \gamma^{t^{f}}\otimes \gamma^{x^{\overline{f}}}\partial_{x}^{\overline{f}}\\
&+\gamma^{y^{f}} \otimes\gamma^{t^{\overline{f}}}\partial_{y}^{f}+\gamma^{t^{f}}\otimes \gamma^{y^{\overline{f}}}\partial_{y}^{\overline{f}}+i\frac{mc}{\hbar}\left[\textbf{I}_{2}\otimes \gamma^{t^{\overline{f}}}+\gamma^{t^{f}}\otimes \textbf{I}_{2}\right].   \label{eq4}
\end{eqnarray}
According to the metric (\ref{metric}), we can factorize the space-time-dependent composite field $\Psi(t,r,R)$ as 
\begin{equation*}
\Psi=e^{-i\omega t}\Tilde{\Psi}(r,R),
\end{equation*}
in which $\omega$ is the relativistic frequency of the considered system, and $R$ and $r$ refer to center of mass motion coordinates and relative motion coordinates, respectively. The center of mass and relative motion coordinates are introduced through \cite{guvendi-mustafa}
\begin{eqnarray}
&r_{\mu}=x_{\mu}^{f}-x_{\mu}^{\overline{f}},\quad R_{\mu}=\frac{x_{\mu}^{f}+x_{\mu}^{\overline{f}}}{2},\quad x_{\mu}^{f}=\frac{1}{2}r_{\mu}+R_{\mu},\nonumber\\
&x_{\mu}^{\overline{f}}=-\frac{1}{2}r_{\mu}+R_{\mu}, \partial_{x_{\mu}}^{f}=\partial_{r_{\mu}}+\frac{1}{2}\partial_{R_{\mu}},\nonumber\\
&\partial_{x_{\mu}}^{\overline{f}}=-\partial_{r_{\mu}}+\frac{1}{2}\partial_{R_{\mu}}, \label{Rr-C}
\end{eqnarray}
for equal masses of two particles. In Eq. (\ref{Rr-C}) it can be seen that $\partial_{x_{\mu}}^{f}+\partial_{x_{\mu}}^{\overline{f}}=\partial_{R_{\mu}}$. This means the evolution of the considered pair, associated with the relativistic frequency, is determined with respect to the proper time $\partial _{R_{t}}$. Here, we will try to explore the relative motion of the considered pair. To acquire this, we need to get rid of the center of motion coordinates. We can consider such a static pair whose center of mass is at rest at the spatial origin. Of course, this requires that the particles must carry opposite momenta with respect to each other. At that rate, we may observe any pairing effect, only. Under this assumption, the resulting equation can be expressed in terms of relative motion coordinates and results in $\hat{\diamondsuit}\Tilde{\Psi}(r_{x},r_{y})=0$ given explicitly by
\begin{equation}
\begin{split}
\left(
\begin{array}{cccc}
\varphi(r)-\tilde{m}&  \hat{\mathcal{D}}^{-} & -\hat{\mathcal{D}}^{-} & 0\\
-\hat{\mathcal{D}}^{+}  & \varphi(r) & 0 & -\hat{\mathcal{D}}^{-}\\
\hat{\mathcal{D}}^{+}   & 0 & \varphi(r) & \hat{\mathcal{D}}^{-}\\
0  & \hat{\mathcal{D}}^{+} & -\hat{\mathcal{D}}^{+} & \varphi(r)+\tilde{m}
\end{array}
\right)\left(
\begin{array}{c}
\psi_{1}(r_{x},r_{y})\\
\psi_{2}(r_{x},r_{y})\\
\psi_{3}(r_{x},r_{y})\\
\psi_{4}(r_{x},r_{y})
\end{array}
\right)=0,\label{ME}
\end{split}
\end{equation}
where $\hat{\mathcal{D}}^{\mp}=\partial_{r_{x}}\mp i\partial_{r_{y}}$, $\varphi(r)=\omega/c-\mathcal{V}(r)$, and $\tilde{m}=2mc/\hbar$. Here, it is clear that the spinor components depend explicitly on the relative motion coordinates, as $\psi_{u}(r_{x},r_{y}), (u=1,2,3,4.)$. In search of a symmetry, we can consider to exploit the angular symmetry in polar space. To map the system into polar coordinates, we can use the spin raising/lowering operators, given by $\hat{\mathcal{D}}_{\pm}=\textrm{e}^{\mp i\phi}\left(\partial_{r}\mp \frac{i}{r}\partial_{\phi}\right)$, only for the transformed spinor components \cite{guvendi2}
\begin{equation*}
\left(
\begin{array}{c}
\psi_{1}(r_{x},r_{y})\\
\psi_{2}(r_{x},r_{y})\\
\psi_{3}(r_{x},r_{y})\\
\psi_{4}(r_{x},r_{y})
\end{array}
\right)\Rightarrow \left(
\begin{array}{c}
\psi_{1}(r)\textrm{e}^{i(s-1)\phi}\\
\psi_{2}(r)\textrm{e}^{is\phi}\\
\psi_{3}(r)\textrm{e}^{is\phi}\\
\psi_{4}(r)\textrm{e}^{i(s+1)\phi}
\end{array}
\right),\label{TC}
\end{equation*}
where $r$ is the relative radial distance between the particles and $s$ is the total spin of the resulting composite system formed by a coupled $f\overline{f}$ pair.

\section{\mdseries{Conditionally exact solutions}} \label{sec:3}

In this section, we derive first a set of coupled equations, and then we show a non-perturbative second-order wave equation for a static and spinless system formed by a $f\overline{f}$ pair interacting through a central (inter-particle) interaction potential $\mathcal{V}(r)$. Then, we will use quasi-exactly solvable methods to analyze the dynamics of the system in question. The matrix equation (\ref{ME}) leads to a set of coupled equations consisting of four first-order differential equations. By adding and subtracting these equations, one can derive following equations set
\begin{eqnarray}
 &\varphi(r)\psi_{+}(r)-\tilde{m}\psi_{-}(r)+4\psi^{'}_{0}(r)=0,\\
 &\varphi(r)\psi_{-}(r)-\tilde{m}\psi_{+}(r)=0,\\
 &\varphi(r)\psi_{0}(r)-\psi^{'}_{+}(r)=0, \label{ES}
\end{eqnarray}
where prime $(^{'})$ indicates derivative with respect to $r$, $\psi_{\pm}(r)=\psi_{1}\pm\psi_{4}$ and $\psi_{0}=\psi_{2}=-\psi_{3}$. The equations (\ref{ES}) can be solved for $\psi_{+}$, and results in the following non-perturbative second-order wave equation
\begin{gather}
\varphi \left( r\right) \,\psi _{+}^{\prime \prime }\left( r\right) -\varphi
^{\prime }\left( r\right) \,\psi _{+}^{\prime }\left( r\right)+\frac{1}{4}\left[ \varphi \left( r\right) ^{3}-\tilde{m}^{2}\varphi \left(
r\right) \right] \psi _{+}\left( r\right) =0,  \label{om1}
\end{gather}%
where $\varphi \left( r\right) =\varpi-\mathcal{V}\left( r\right) $, $\varpi=\omega /c$, $\mathcal{V}\left( r\right) =A\exp \left( -r/\tilde{\lambda}%
\right) $, $\tilde{m}=2mc/\hbar =2/\lambda $,  and $\tilde{\lambda}=\lambda d
$. We have also to use the change of variables $x=\mathcal{V}\left( r\right) /\varpi\in [0,A/\varpi]$ to obtain
\begin{gather}
x^{2}\psi _{+}^{\prime \prime }\left( x\right) +\frac{x}{1-x}\,\psi
_{+}^{\prime }\left( x\right)+\frac{1}{4}\left[ -\alpha^2 x^{2}+2\alpha^{2}x-\beta^2\right] \psi _{+}\left(
x\right) =0,  \label{om2}
\end{gather}%
where $\alpha =i\tilde{\lambda}\varpi$ and $\beta=d\sqrt{4-\lambda^2\varpi^2}$ are used here. Let us use the substitution
\begin{equation}
\psi _{+}\left( x\right) =\mathit{N}\,x^{\beta /2}\,e^{-\alpha x/2}H\left( x\right) ,  \label{om21}
\end{equation}
to obtain
\begin{gather}
x\left( 1-x\right) H^{\prime \prime }\left( x\right)+\left[ \alpha x^{2}-\left( \beta +\alpha \right) x+\left( \beta +1\right) %
\right] H^{\prime }\left( x\right)+\frac{1}{2}\left[ \alpha \left( \beta -\alpha \right) x+\zeta \right]
H\left( x\right) =0,  \label{om5}
\end{gather}
where 
\begin{equation}
\zeta =(1-\alpha)(\beta-\alpha) .  \label{om6}
\end{equation}
Next we use the power series expansion 
\begin{equation}
H\left( x\right) =\sum_{j=0}^{\infty }C_{j}\,x^{j+\sigma },  \label{om7}
\end{equation}
in (\ref{om5}) to obtain
\begin{gather}
\sum_{j=0}^{\infty }C_{j}\left[ \alpha \left( j+\sigma \right) +\frac{
\alpha }{2}\left( \beta -\alpha \right) \right] \,x^{j+\sigma +1}  \nonumber \\
+\sum_{j=0}^{\infty }C_{j}\left[ \frac{\zeta}{2}-\left( j+\sigma \right) \left( j+\sigma-1\right) -\left( j+\sigma \right) \left( \beta +\alpha \right) \right] \,x^{j+\sigma } \nonumber \\
 +\sum_{j=0}^{\infty }C_{j}\,\left[ \left( j+\sigma \right) \left( j+\sigma
-1\right) +\left( j+\sigma \right) \left( \beta +1\right) \right]
\,\,x^{j+\sigma -1}=0.  \label{om8}
\end{gather}
This would, in turn, imply that
\begin{gather}
\sum_{j=0}^{\infty }\left\{ C_{j+2}\left[ \left( j+\sigma +1\right) \left(
j+\sigma +2\right) +\left( j+\sigma +2\right) \left( \beta +1\right) \right]
\right.   \nonumber \\
\left. -C_{j+1}\left[ \left( j+\sigma \right) \left( j+\sigma +1\right)
+\left( j+\sigma +1\right) \left( \beta +\alpha \right) -\frac{\zeta }{2}%
\right] \right.   \nonumber \\
\left. +C_{j}\left[ \alpha \left( j+\sigma \right) +\frac{\alpha }{2}\left(
\beta -\alpha \right) \right] \right\} \,x^{j+\sigma +1}=0,  \label{om9}
\end{gather}
provided that $\sigma =0$, and
\begin{equation}
C_{1}=-\frac{\zeta }{2\left( \beta +1\right)} C_{0}. \label{om10}
\end{equation}%
We may now obtain the three terms recursion relation
\begin{gather}
C_{j+2}\left[ \left( j+1\right) \left( j+2\right) +\left( j+2\right)
\left( \beta +1\right) \right]+C_{j+1}\left[\frac{\zeta }{2}- j\left( j+1\right)  
-\left( j+1\right) \left( \beta +\alpha\right) \right] \nonumber \\
+C_{j}\left[ \alpha j+\frac{\alpha }{2}\left( \beta -\alpha \right) \right]=0. \label{om11}
\end{gather}

\begin{figure}
	\centering 
	\includegraphics[width=1\textwidth]{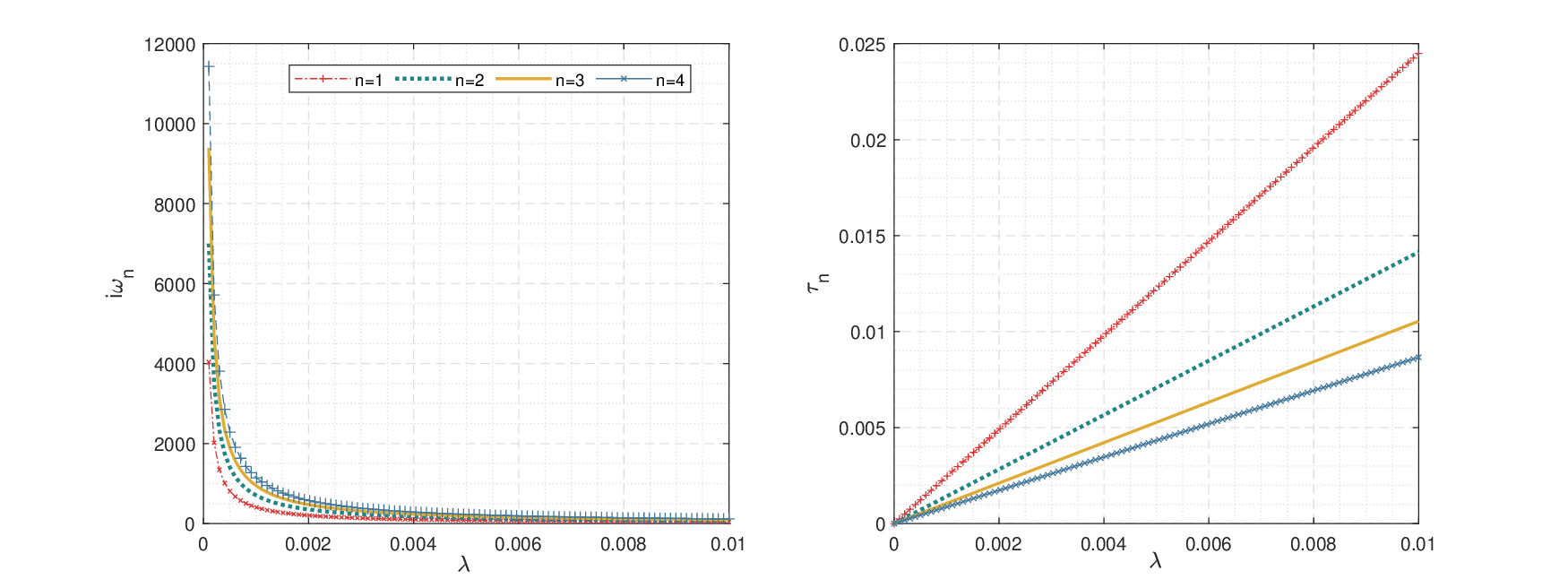}	
	\caption{The frequency $i\omega_{n}$ and corresponding decay time ($\tau_{n}$) vary with the Compton wavelength $\lambda$. Here we take $c=1$.} 
	\label{fig:2}
\end{figure}

We may now truncate the power series (\ref{om7}) to a polynomial of order $n+1\geq1$ by imposing the conditions that $\forall j=n$ we have $C_{n+2}=0$, $C_{n+1}\neq 0$, and $C_{n}\neq 0$. The first condition $C_{n+2}=0$ truncates the power series to a polynomial of order $n+1$. However,  since $C_{n+1}\neq 0$, and $C_{n}\neq 0$,  one may use the conditions that the coefficients of $C_{n+1}\neq 0$ and $C_{n}\neq 0$ identically vanish to facilitate conditionally exact solvability of the problem (c.f., e.g., \cite{o1,o2}). That is, for $C_{n}\neq 0$ we have
\begin{equation}
\alpha n+\frac{\alpha }{2}\left( \beta -\alpha
\right) =0\Longrightarrow  \beta=\alpha -2n,  \label{om12}
\end{equation}
and for $C_{n+1}\neq 0$ we obtain
\begin{eqnarray}
n\left( n+1\right) +\left( n+1\right) \left(
\beta +\alpha \right) =\frac{\zeta }{2},  \label{om13}
\end{eqnarray}
to imply that 
\begin{equation}
\frac{(1-\alpha)(\beta-\alpha)}{2}=n\left( n+1\right) +\left( n+1\right) \left(
\beta +\alpha \right). \label{om14}
\end{equation}
We may now use (\ref{om12}) and (\ref{om14}) to obtain
\begin{equation}
    \alpha=\frac{n^2}{n+2} \Rightarrow i\omega=\frac{c}{\lambda d}\left[\frac{n^2}{n+2}\right]. \label{om15}
\end{equation}
Moreover, $\beta=\sqrt{4d^2+\alpha^2}$ and $\beta=\alpha -2n$, one obtains
\begin{equation}
d=d_n=\pm\frac{2n}{\sqrt{2n+4}} \Rightarrow d_n=\frac{2n}{\sqrt{2n+4}}>0. \label{om16}
\end{equation}
That is, for every $n$ we have a different $d=d_n$ value.  This relation would therefore identify a correlation between our parameter $d$ and the quantum number $n$.  Consequently, the results in (\ref{om12}), (\ref{om14}), and (\ref{om15}) would yield
\begin{equation}
i\omega_{n}=\frac{c}{\lambda d_n}\left[\frac{n^2}{n+2}\right]=\frac{c}{\lambda}\left[\frac{n}{\sqrt{2n+4}}\right]\Rightarrow \hbar \omega_{n}=-i\,mc^2\left[\frac{n}{\sqrt{2n+4}}\right].\label{om17}
\end{equation}
Only under such conditions that our power series is truncated to a polynomial of order $n+1\geq1$.  Therefore, the condition that the coefficients of $C_{n+1}\neq 0$ and $C_n\neq0$ vanish identically would  manifestly facilitates conditionally exact solvability of the problem at hand. Notably, such a conditionally exact solution puts some parametric constraint (in our case, the allowed values of $d=d_n;\,n>0$, loosing $n=0$ state in the process of conditionally exact solvability, therefore), which is, in fact, a usual price one has sometimes to pay when conditional exact solvability is involved. Very recently, similar conditionally exact solvability for such three terms recursion relations is suggested for the confluent \cite{o1} and the biconfluent \cite{o2} Heun type functions/series solutions. The frequency expression (\ref{om17}) allows us to determine the decay time ($\tau_{n}=1/|\omega_{\mathfrak{Im}_{n}}|$) of the damped modes
\begin{eqnarray}
\tau_{n}=\frac{\lambda}{c}\frac{\sqrt{2n+4}}{n},   \label{DT}
\end{eqnarray}
provided $n\geq1$ (note that $\Psi\propto \exp(-i\omega t)$ ). The relationship between decay time and the Compton wavelength $\lambda$ is evident in Fig. \ref{fig:2}. Additionally, it is apparent that the decay time of this composite system can be very long if the Compton wavelength is very long. However, each state decays faster for small $\lambda$ values.  

\section{\mdseries{Summary and discussions}}\label{sec:4}

This study explores the relative motion of a $f\overline{f}$ pair in ($2+1$)-dimensional flat space-time, interacting via an exponentially decaying inter-particle interaction potential. We seek analytical solutions to the fully-covariant one-time two-body Dirac equation derived from quantum electrodynamics using the action principle. For such a spinless composite system, we derive a non-perturbative second-order wave equation and find its solution by establishing quantization condition for the formation of such pairs. This approach enables us to determine relativistic frequency modes. The frequency modes $\omega_{n}$ are found in purely imaginary form, representing the time evolution of the composite field \(\Psi \propto \exp(-i\omega_{n}t)\). The explicit form of the frequency spectra is given by: 
$\mathfrak{Im}(\omega)_{n}=-i\,\frac{c}{\lambda}\frac{n}{\sqrt{2n+4}}$. Here, $\lambda$ denotes Compton wavelength for the fermion (or its antiparticle, the antifermion), $c$ is the speed of light, and $n$ is the overtone number ($n=1,2,\cdots$). Moreover, these states cannot be steady states and decay over time. Thus, the time evolution of the system depends explicitly on the Compton wavelength but is controlled also by the relationship between the scaling factor and the overtone number $n$. Notably, the time evolution is independent of the strength of the interaction. Our results show that frequency modes always have a negative signature, indicating that the corresponding quantum states decay over time, with decay time ($\tau_{n}$) depending the Compton wavelength and the scaling since $\tau_{n}=\frac{\lambda}{c}\frac{\sqrt{2n+4}}{n}$. The dependence of the decay time of these damped modes is illustrated in Fig. \ref{fig:2}, showing that the decay time of such $f\overline{f}$ pairs is dominated by the $n=1$ state, especially for long $\lambda$ (or for small rest mass). These results suggest also that all physically possible quantum states decay faster if $\lambda$ is small. 

On the other hand, we may feasibly extend our findings to $f\overline{f}$ pairs within condensed matter systems, utilizing the Fermi speed rather than the speed of light in vacuum, could prove highly beneficial. Given that the Fermi speed $(v_{F})$ typically falls considerably below the speed of light ($c$). Our findings suggest that fermionic states within condensed matter environments can exist for longer duration compared to their vacuum counterparts. This adaptation can be useful for modeling and comprehending phenomena in condensed matter physics, where pair behavior is of course influenced by also environmental and material characteristics. Consequently, in principle, we can apply our findings to $f\overline{f}$ pairs in condensed matter systems by substituting the speed of light in vacuum with the Fermi velocity $(v_{F}\approx c/300)$. Essentially, this implies that the decay time of such interacting $f\overline{f}$ pairs in condensed matter mediums can be approximately $3\times 10^{2}$ times longer. Our result given by Eq. (\ref{DT}) may be applied to electron-hole pairs in two-dimensional materials without loss of generality. In this context, the result (\ref{DT}) becomes
\begin{eqnarray}
\tau_{1}=\sqrt{6}\,\frac{\lambda}{v_{F}},   \label{DT-1}
\end{eqnarray}
when \( n=1 \). This suggests a decay time of approximately \( \sim 10^{-17} \) seconds for electron-hole systems in the S-state within monolayer materials, even without accounting for specific material properties and thermal effects (see also \cite{exc}). This finding aligns with the results obtained for excitons in certain two-dimensional materials \cite{exc}. Furthermore, our model could be very useful in explaining the dependence of the decay time of excitonic states \cite{exc} on the effective dielectric constant of monolayer materials, which could be explored in future research.

Exploring how $f\overline{f}$ pairs interact under exponentially decaying potentials is a fascinating area in physics \cite{book1,book2} (see also \cite{Sol}). These potentials appear in different physical scenarios, reflecting a range of interactions, such as those mediated by virtual particles in quantum field theory. Understanding the dynamics of $f\overline{f}$ pairs in these conditions could enhance our grasp of particle physics in both flat and curved spaces.


\section*{Data Availability}

No new data were generated or analyzed in this study.

\section*{Conflict of Interests}

Authors declare no conflict of interests.

\section*{Funding Statement}

No fund has received for this research.

\end{document}